\documentclass[preprint2]{aastex}

\shorttitle{SPIRAL STRUCTURE IN DWARF GALAXIES}
\shortauthors{Graham, Jerjen, Guzm\'an}

\begin{document}

\title{{\it HST} DETECTION OF SPIRAL STRUCTURE IN TWO COMA CLUSTER DWARF GALAXIES\altaffilmark{1}}

% One preprint or reprint of each refereed publication based on HST
% archival research must be sent to the following address:
% Librarian - Space Telescope Science Institute - 3700 San Martin Drive
% - Baltimore, Maryland 21218 USA

\author{Alister W.\ Graham}
\affil{Department of Astronomy, University of Florida, P.O.\ Box 112055, Gainesville, FL 32611, USA}
\email{Graham@astro.ufl.edu}
%\and
\author{Helmut Jerjen}
\affil{Research School of Astronomy and Astrophysics, Australian National University, Private Bag, Weston Creek PO, ACT 2611, Canberra, Australia}
\and
\author{Rafael Guzm\'an}
\affil{Department of Astronomy, University of Florida, P.O.\ Box 112055, Gainesville, FL 32611, USA}

\altaffiltext{1}{Based on observations made with the NASA/ESA {\it 
Hubble Space Telescope}, obtained at the Space Telescope Science
Institute, which is operated by the Association of Universities for
Research in Astronomy, Inc., under NASA contract NAS 5-26555.}

\begin{abstract}
We report the discovery of spiral-like structure in Hubble Space Telescope 
images of two dwarf galaxies (GMP~3292 and GMP~3629) belonging to the 
Coma cluster.  GMP~3629 is the faintest such galaxy 
detected in a cluster environment, and it is the first such galaxy observed 
in the dense Coma cluster.  The large bulge and the faintness of the 
broad spiral-like 
pattern in GMP~3629 suggests that its disk may have been largely depleted. 
We may therefore have found an example of the ``missing link'' in 
theories of galaxy evolution which have predicted that dwarf spiral 
galaxies, particularly in clusters, evolve into dwarf elliptical galaxies. 
\end{abstract}


\section{Introduction}

Evidence for embedded, geometrically-flat, stellar disks has been 
found in a steadily increasing number of objects that were once 
regarded as purely elliptical (E) galaxies 
% Burstein 1979; Simien \& de Vaucouleurs 1986; 
(e.g., Capaccioli 1987, his section 5; 
Rix \& White 1990; Vader \& Vigroux 1991; Nieto et al.\ 1992; 
Cinzano \& van der Marel 1994; Jorgensen \& Franx 1994; 
Sahu, Pandey, \& Kembhavi 1996). 
% Some Authors have even gone so far as to speculate that {\it all} 
% E galaxies may contain large-scale disks (Burstein 2001; 
% Saglia et al.\1997) because of departures from the 
% $R^{1/4}$ model. 
% Binggeli \& Cameron (1991) actually suggested that most nucleated, 
% bright dwarfs may contain disks. 
It has thus become a somewhat common trend for E 
galaxies to be reclassified as lenticular (S0) galaxies.  
This can happen when unsharp masking 
% (e.g., Buta \& Crocker 1993; Erwin \& Sparke 2002) 
of a galaxy's image reveals features indicative 
of a disk, such as bars and/or spiral patterns, 
or after a closer examination of a galaxy's 
surface brightness profile reveals multiple component structure
(e.g., Scorza et al.\ 1998).  
% (not simply departures from the $R^{1/4}$ model, Burstein et al.\ 2001)
It can, and has, also occurred after the inspection of kinematical data 
reveals significant rotation and/or bar-like dynamical behavior 
(e.g., Carter 1987; Nieto, Capaccioli, \& Held 1988; 
Capaccioli \& Longo 1994; Scorza \& Bender 1995; Graham et al.\ 
1998; Rix, Carollo, \& Freeman 1999). 

Similarly, the existence of previously undetected disks in dwarf
elliptical (dE) galaxies are now being realized.  Although it has been
known for nearly twenty years that some dE-like galaxies do have
disk-like morphologies (dS0; Sandage \& Binggeli 1984; Binggeli \& Cameron
1991), what is new --- in addition to the rising number of disk
detections --- is that some dE-like galaxies actually have (stellar)
spiral structures in their disk.  After the initial surprise
announcement of a tightly-wound, two-armed spiral structure in the dE
galaxy IC~3328 (Jerjen, Kalnajs, \& Binggeli 2000b), Jerjen, Kalnajs
\& Binggeli (2001) and Barazza, Binggeli, \& Jerjen (2002) reported
previously undetected spiral structure and bars (i.e., disks) in four
% 
% IC3328: Jerjen et al. 2000
% IC3328, IC0783: Jerjen et al. 2001
% IC 0783, IC 3349, NGC 4431, IC 3468: Barazza et al. 2002
% 
more Virgo cluster galaxies (2 dS0, 1 dE, and 1 low-luminosity E).  
De Rijcke et al.\ (2003) have additionally presented photometric and 
kinematic evidence for disks, and in one case spiral arms, in two 
edge-on Fornax cluster dS0 galaxies.  
% HI gas is additionally detected in about $\sim$15\% of 
% early-type dwarf galaxies (e.g., Conselice et al.\ 2003). 
The data to date suggests that up to 20\% of bright early-type 
dwarf galaxies in clusters may have disks. 
% Recent kinematical measurements of dS0 galaxies 
% are starting to reveal that these galaxies can indeed possess significant 
% rotational energy (e.g., UGC~7436 
% described in Ferguson \& Binggeli 1994; De Rijcke et al.\ 2001a,b; 
% Geha et al.\ 2002; Pedraz et al.\ 2002; Simien \& Prugniel 2002; 
% Thomas et al.\ 2002; Zeilinger et al.\ 2002)
%% \footnote{It is noted that at least one dE galaxy (the E1 galaxy FS~76; 
%% De Rijcke et al.\ 2001a) {\it is} mildly flattened by rotation rather 
%% than by velocity anisotropy.  However, rotation is not the dominant 
%% kinetic energy term.}\\
We report here on the first ever detection of spiral-like structure in 
two dwarf, early-type galaxies residing in the densest cluster environment 
studied so far, namely the Coma cluster.  
% The commonality of embedded disks in both bright (classical) 
% and faint (dwarf) lenticular galaxies suggests, but is certainly not 
% conclusive, that similar formation mechanisms may have been operational. 

The presence, or at least detection, of (stellar) spiral patterns 
in dwarf galaxies is a particularly rare phenomenon. 
% late-type dS galaxies have also been reported 
% (Matthews \& Gallagher 1997). McGaugh (1994), de Blok et al.\ (1995)
Although dwarf versions of Sm and 
Irr galaxies have been known for a long time (e.g., van den Bergh 
1960), referring to the early-type Sa-Sc spiral galaxies,  
Ferguson \& Sandage (1991) wrote that ``dwarf spiral galaxies 
do not appear to exist'' (see also Sandage \& Binggeli 1984 and Sandage, 
Binggeli \& Tammann 1985).  They are a rare species; indeed, their very 
existence was only recognized a few years ago (Schombert et al.\ 1995). 
Even then, Schombert et al.\ concluded that dwarf spiral 
galaxies only exist in the field.  The harsh environment within 
a galaxy cluster --- due to galaxy mergers, with each other or 
the intracluster medium, and/or strong gravitational tidal 
interactions --- is commonly thought to have led to the destruction 
of the delicate spiral patterns in dwarf galaxies. 
A comparison of the number of such objects in low- and high-density 
environments may shed light on the nature of their existence. 

By searching for signs of apparent spiral structure and/or bars in the optical 
images from the sample of 18 dE galaxy candidates presented in Graham 
\& Guzm\'an (2003), we explore here which galaxies may have embedded 
stellar disks.  
The galaxy selection criteria is described in the following section, 
as is the image reduction process and analysis. Section 3 provides 
a brief quantitative analysis of the disks, and Section 4  
discusses possible evolutionary scenarios for dwarf disk galaxies
in clusters.  

We take Coma to be at a distance of 100 Mpc and 
use $H_0$=70 km s$^{-1}$ Mpc$^{-1}$, 0.1$\arcsec$ therefore
corresponds to 47 pc.

\section{Galaxy Sample and Image Analysis}

Galaxies meeting the following conditions --- discussed at 
length in Matcovi\'c \& Guzm\'an (2003, in prep) --- were 
selected from the Coma cluster field catalog (Godwin, 
Metcalfe \& Peach 1983; hereafter GMP).  All galaxies have 
positions within the central $20\arcmin \times 20\arcmin$ of the 
Coma cluster;  $-17.5 < M_B < -14.5$;  
$0.2 < (U-B) < 0.6$ and $1.3 < (B-R) < 1.5$; available {\it HST} WFPC2 
images and recessional velocities between 4,000 and 10,000 km s$^{-1}$.  
The spectral analysis and recessional velocity derivation 
is also provided in Matcovi\'c \& Guzm\'an (2003, in prep). 
The above requirements were expected to result in the selection 
of Coma cluster dwarf elliptical galaxies, and we obtained 18 
such candidates.  With the exception of GMP~2960, there were no 
pre-existing morphological type classifications for these galaxies. 
GMP~2960 (PGC 44707; Paturel et al.\ 1989) 
% (RB 074; Rood \& Baum 1967) 
is classified in NED as an S0 galaxy, and 
according to the type-specific luminosity functions derived from 
three clusters (Jerjen \& Tammann 1997) we conclude that 
GMP~2960 is either a low-luminosity S0 or a bright dS0 galaxy.

The reduction process of the HST images is described in 
Graham \& Guzm\'an (2003).  Briefly, we used the \textsc{iraf} task
\texttt{crrej} to combine the {\it HST}-pipelined exposures, 
which we then further cleaned of cosmic rays using 
\texttt{LACOS} (L.A.COSMIC, van Dokkum 2001).  Due to the stellar 
halos of nearby galaxies, we used the wavelet decomposition method 
of Vikhlinin et al.\ (1998) to simultaneously subtract this non-uniform 
light and the sky background.  Foreground stars and 
over-lapping background 
galaxies were searched for, and masked out, before we performed 
any image analysis or surface brightness fitting.

In order to search for non-symmetric structures in the dwarf galaxy 
images, we subtracted the axisymmetric component of the galaxy light 
% (i.e., the $I_0(r)$ term in their equation 1) 
(see Jerjen et al.\ 2000b), leaving a 
``residual image''. 
Following Barazza et al.\ (2002) and De Rijcke et al.\ (2003), 
we have additionally used an unsharp masking technique to verify the 
presence of features such as bars or spiral arms, which 
would indicate the presence of a flattened stellar disk.  Although 
the majority of galaxies showed no sign of non-axisymmetric structure, 
two galaxies (GMP~3292 and GMP~3629) were found to possess flocculent 
spiral arms (Figures 1-2).  Their basic properties are given in Table 1. 

From our previous analysis of the radial light-profiles 
(Graham \& Guzm\'an 2003), we had already identified 
GMP~3292 as a likely bulge/disk system due to a clear break 
in its surface brightness profile marking the bulge/disk transition.  
With regard to 
GMP~3629, we had remarked upon the possibility of an outer 
disk --- not dominating until radii greater than $\sim 10\arcsec$ 
(4.7 kpc) --- 
but we could not and did not confirm this due to the low 
surface brightness levels at these outer radii. We can however now 
confirm that both of these galaxies possess stellar disks as 
indicated by the presence of a spiral pattern. 
Using the velocity catalog of Edwards et al.\ (2002),
Guti\'errez et al.\ (2003) derive a mean recessional velocity
of 6862 km s$^{-1}$, and a velocity dispersion of 1273$\pm$145 km
s$^{-1}$, for Coma dwarf galaxies fainter than $M_B=-17.5$ mag.
The recessional velocities of GMP~3292 (4955 km s$^{-1}$) and 
GMP~3629 (5219 km s$^{-1}$) are therefore consistent with membership 
in the Coma cluster.  
% We do however not the possibility that these galaxies may be 
% foreground galaxies.  
Although, these velocities are 1.5 and 1.3 sigma from the mean 
cluster value, one may actually expect to find such an offset given 
the ``infalling group'' feeding mechanism for clusters 
(see, e.g., Zabludoff \& Franx 1993; Conselice, Gallagher, \& Wyse 
2001; Drinkwater, Gregg, \& Colless 2001a). 
% Mihos 2003 - r^(1/4) rubbish). 

From the full sample of 18 dwarf galaxies, 
Graham \& Guzm\'an (2003) had additionally found that 
neither GMP~2960 nor GMP~3486 could be described with a single 
S\'ersic model; that is, 
their surface brightness profiles suggested the presence of 
more than one component (aside from nucleation). 
However these two galaxies display no evidence of spiral, or in fact 
any asymmetric, structure in their residual images.  GMP~3486 may therefore 
be, like the previously mentioned classification for GMP~2960, a 
lenticular galaxy with a
large-scale disk displaying no obvious spiral structure. 
Neither the light-profile analysis in Graham \& Guzm\'an (2003) 
nor the present residual image analysis provide evidence to suggest that the 
remaining 14 galaxies are anything but nucleated dwarf elliptical 
galaxies.

\section{Quantitative Results}

The inwardly extrapolated exponential disk in GMP~3292 has a central 
surface brightness of $\mu_{0,F606W}$=20.68 mag arcsec$^{-2}$ 
(Graham \& Guzm\'an, their table 2). 
Correcting this value for Galactic extinction ($-$0.02 mag; 
Schlegel, Finkbeiner, \& Davis 1998), $(1+z)^4$ redshift 
dimming ($-$0.10 mag), 
and $K$-correction (0.02 mag; Poggianti 1997) gives a 
value of 20.58 mag arcsec$^{-2}$. 
Assuming a $B-${\it F606W} color of 1.08 
(Fukugita, Shimasaku, \& Ichikawa 1994) yields\footnote{A small 
but unknown inclination correction for dust may be required; the 
outer galaxy isophotes have an ellipticity of only 0.20.} 
$\mu_{0,B}$=21.66 mag arcsec$^{-2}$, very close indeed to the 
canonical Freeman (1970) value of 21.65 $B$-mag arcsec$^{-2}$.  
The disk scale-length is 2.12$\arcsec$, which translates to $\sim$1 kpc. 
This is at the small end of the range from 1.0 to 2.5 kpc found in 
Schombert et al.'s (1995) sample of dwarf spiral 
galaxies\footnote{The disk scale-length range given in Schombert 
et al.\ (1995) increases to 1.2-3.0 kpc when using $H_0$=70 
km s$^{-1}$ Mpc$^{-1}$.}. 
The total apparent galaxy magnitude is 16.74 {\it F606W}-mag 
(Graham \& Guzm\'an 2003, their table 2).  
Assuming a distance modulus of 35.0 gives a corrected absolute 
$B$-band magnitude of -17.28 $B$-mag. 
% 
% The central disk surface brightness in GMP~3292 is much 
% brighter than the value of 23-24.0 $B$-mag arcsec$^{-2}$ for the 
% dwarf spiral galaxies discussed in Schombert et al.\ (1995) ... 
%% Schombert et al.\ were referring to the *galaxy* (not *disk*) central SB. 
% 
Based on the luminosity functions of individual Hubble types 
(Jerjen \& Tammann 1997, their Fig.3), 
GMP~3292 is likely to be a small late-type (Sc-Sm) spiral galaxy. 
% Although its bulge-to-disk luminosity ratio 
% (log[B/D]=-0.50) is more typical of an early-type disk galaxy, 
% this ratio is consistent with values found for 
% intermediate-type spiral galaxies (c.f.\ Graham 2001, his Figure 15); 
The placement of its disk in the $\mu_0-log(h)$ diagram 
(Figure~\ref{fig3}) is also consistent with this assignment, 
although it should be kept in mind that this region of 
parameter space containing small faint disks is far from fully 
explored.

The second galaxy, 
GMP~3629, is modeled here in an identical fashion to Graham \& 
Guzm\'an's treatment of GMP~3292.   A Moffat function was fitted to 
nearby stars  and/or globular clusters on the {\it HST} WFPC chip 
containing the image of GMP~3629.  The best-fitting Moffat function 
was then used to convolve the central Gaussian, S\'ersic $R^{1/n}$, and 
exponential models which were simultaneously 
fitted to GMP~3629's nuclear star cluster, bulge, and disk respectively
(see Figure~\ref{fig4}).  The resultant central disk surface brightness of 
25.58 {\it F606W}-mag arcsec$^{-2}$ translates into a corrected 
$B$-band value of 26.56 mag 
arcsec$^{-2}$ --- 5 mag arcsec$^{-2}$ fainter than the Freeman value, 
suggestive of depletion. 
This value is however somewhat poorly 
constrained and a deeper exposure would be of great value for better 
quantifying the disk.

The total apparent {\it F606W} magnitude of GMP~3629 
is 18.02 mag --- derived from the best-fitting star-cluster/bulge/disk 
models extrapolated to infinity.  
This agrees well with the apparent {\it F606W} magnitude of 18.18 mag 
obtained in Graham \& Guzm\'an (2003)\footnote{Graham \& Guzm\'an (2003) 
derived the total galaxy magnitude for GMP~3629 from an $R^{1/n}$-bulge 
(plus star-cluster) fit to the surface brightness profile; no 
disk was fitted.}  and the 
apparent {\it blue} magnitude 19.03 $B$-mag obtained by Godwin et al.\ 
(1983) within the isophote $b_{26.5}$.  The value 18.02 $F606W$-mag 
translates into a corrected absolute blue magnitude $M_B$=-16.00 mag, 
which happens to mark the boundary between bright and faint dwarf galaxies 
(Ferguson \& Binggeli 1994). 
% dwarf galaxies brighter than $M_B$=-16 are 
% almost exclusively found in high density environments. 
This is also the limiting magnitude\footnote{Sandage et al.\ (1985) used 
a Virgo cluster distance modulus of $m-M$=31.7.} where Sandage et al.\ 
(1985) claimed fainter (Virgo cluster) spiral galaxies did not 
exist (see also Ferguson \& Sandage 1991).  Assuming a $B-V$ color of 0.9, 
% (ref? Faber et al 1997, bright Es) 
GMP~3629's magnitude is equivalent to the magnitude of three of the six 
``dwarf spiral'' galaxies presented in Schombert et al.\ (1995). 
Unlike GMP~3292, GMP~3629 is faint enough to meet Schombert et al.'s 
selection criteria for``dwarf spiral'' galaxies.  

In addition to having faint magnitudes 
($M_B \gtrsim -16$ mag), 
and faint central surface brightness values (23-24 $B$-mag arcsec$^{-2}$), 
Schombert et al.\ (1995) identified 
additional characteristics of their dwarf spiral galaxies. 
They found that the optical radii were small, with $R_{26}<$ 5-6 kpc
(see their table 1)\footnote{These values increase to 6-7 kpc when
using $H_0$=70 km s$^{-1}$ Mpc$^{-1}$}.  The GMP catalog gives an
isophotal radius for GMP~3629 of $b_{26.5}=10.1\arcsec$ (4.7 kpc);
although the $R_{26}$ radius is expected to be slightly larger, this
is nonetheless a small galaxy.  Schombert et al.\ also remarked on the
presence of gas and dust.  More specifically, they observed a
flocculent nature to the spiral arms {\it and} they measured ``double
horned'' shapes in the HI line profiles of their galaxies.  Their
dwarf spiral galaxies therefore contain significant gas and their
disks do rotate.  Unfortunately we have no rotational data for GMP~3629.
As for the presence of gas and dust, there is no obvious signs of this
either.  However, if GMP~3629 has been somewhat harassed by the
cluster environment, then we would indeed expect to find that the gas
and dust have been stripped away from this galaxy's disk, or 
to have sank to the center of the galaxy where new stars may
have formed.  We may also 
expect the stars in the original disk to have been heated up to create 
the bulge-like structure we see, and the spiral arms to have
simultaneously broadened to their present state. 
It does therefore seem plausible that GMP~3629 may be one of the
first dS galaxies detected in a cluster, although, undergoing a
metamorphism.  Technically, it is no longer a dwarf spiral galaxy, but
it is by no means (yet?) a dwarf elliptical galaxy either.  From the
decomposition of the light-profile (Figure~\ref{fig4}), the disk has
only 38\% the luminosity of the bulge --- a very low value which may
be indicative that much of the disk has either been removed from the
galaxy or redistributed.  The disk-to-bulge luminosity ratios for the
early-type dS galaxies in Schombert et al.\ (1995) were such that
their disks were, relative to their bulges, some $\sim$200-600\% more
luminous.

\section{Discussion}

The nomenclature for dwarf galaxies identifies a number of 
different species.  At least structurally, the dwarf elliptical 
galaxies ($-13\lesssim M_B\lesssim -18$) are now known to be 
the low luminosity extension of ordinary, bright ($M_B\lesssim -18$) 
elliptical galaxies (Jerjen \& Binggeli 1997; Jerjen, Binggeli, \& 
Freeman 2000a; Graham \& Guzm\'an 2003)\footnote{The term 
``dwarf elliptical'' should not to be confused with the 
rare class of ``compact elliptical'' galaxy (de Vaucouleurs 1961), 
whose very existence has recently been questioned (Graham 2002).}.  
The gas-deficient\footnote{Gas has, however, been detected 
{\it surrounding} the optical component of dSph galaxies 
(e.g., Carignan 1999; Blitz \& Robishaw 2000).}
(e.g., Skillman \& Bender 1995; Young 2000), low surface brightness 
($\mu_{0}\gtrsim 23$ $B$-mag arcsec$^{-2}$) dwarf spheroidal (dSph) 
galaxies are fainter still ($M_B\lesssim -13$, Grebel 
2001)\footnote{For a long time dSph galaxies were largely only 
observed in the Local Group (e.g., Da Costa 1998; Mateo 1998; 
Armandroff, Jacoby, \& Davies 1999; Caldwell 1999; van den 
Bergh 1999, 2000), but have now been detected in other groups 
(e.g., Jerjen et al.\ 2000a; Zabludoff \& Mulchaey 2000; 
Carrasco et al.\ 2001) 
and also in the Virgo and Fornax clusters (Phillipps et al.\ 1998; 
Drinkwater et al.\ 2001b; Hilker, Mieske, \& Infante 2003).}.  
Their location in the magnitude--central surface brightness 
diagram reveals a continuous extension with the dE and E galaxies 
% , albeit with an increase of scatter --- possibly due to the use
% inwardly extrapolated exponential models, and certainly due to 
% the broader range of colors these objects exhibit (Conselice 2002, 
% his fig1) ---, 
and hence the division of galaxy classes is somewhat 
artificial.  Reflecting this is the fact that many 
Authors don't even bother to make the distinction between dE and dSph 
galaxies; although, the latter may have a greater range of 
formation histories (e.g., Conselice 2002) 
They are also generally recognized as increasingly 
dark matter dominated (e.g., Carignan \& Freeman 1988; Irwin \& 
Hatzidimitriou 1995; Mateo 1997; Kleyna et al.\ 2002), but see 
Miligrom (1995), Klessen \& Kroupa (1998), 
Klessen \& Zhao (2002) and, in the case of Ursa Minor, 
Gomez-Flechoso \& Martinez-Delgado (2003).  

The gas-rich, clumpy (both optically and in HI gas) dwarf irregular 
(dIrr) galaxies may be the progenitors of some dSph, dE, and dS0
galaxies.  Although dIrr galaxies are considered to be 
rotating disk galaxies, such a morphological transformation may occur 
via galaxy merging (Toomre \& Toomre 1972; Toomre 1974; 
% Schweizer 1982; 
Barnes \& Hernquist 1996; Bekki 1998; Burkert \& Naab 2003), 
or via the somewhat less severe ram-pressure (and turbulent and 
viscous) stripping of gas from a galaxy as it moves through the 
intracluster medium (e.g., Gunn \& Gott 
1972; Nulsen 1982; Lin \& Faber 1983; Cayatte et al.\ 1994; Sofue 1994; 
van den Bergh 1994; Quilis, Moore, \& Bower 2000; Fujita 2001; Toniazzo 
\& Schindler 2001; Grebel, Gallagher, \& Harbeck 2003; Lee, McHall, \& 
Richer 2003).  Although, this latter scenario can not be applicable to
the brighter dEs, which are more massive than the dIrr galaxies 
(Bothun et al.\ 1986). 

Another scenario is ``galaxy harassment'' (Moore et al.\ 1996; Moore,
Lake, \& Katz 1998; Mao \& Mo 1998; Mayer et al.\ 2001) which not 
only considers the perturbing influence of the entire cluster's 
gravitational field  
(e.g., Byrd \& Valtonen 1990) but also tidal effects from repeated, 
fast flybys of massive galaxies. 
Tidal heating from galaxy-galaxy and galaxy-cluster interactions 
is expected to thicken a galaxy's disk (e.g., T\'oth \& Ostriker 1992) 
and suppress spiral 
features while ram-pressure stripping can remove 
the gas and prevent further star formation --- producing S0 and dS0 
galaxies which are characterized by their thick featureless disks. 
Additionally, ``galaxy threshing'', an extreme example of tidal forces
(Bekki, Couch, \& Drinkwater 2001a), has been invoked to explain the
formation of M32-like objects (Bekki et al.\ 2001b; Graham 2002), and
the recently discovered ``ultra-compact'' dwarf galaxies (Drinkwater et
al.\ 2003) --- purportedly the nuclear remnants of nucleated dE/dSph
galaxies which have been severely tidally stripped,
Lastly, ``galaxy starvation'' --- the removal of halo gas, as opposed to 
disk gas, from spiral galaxies falling into a galaxy cluster --- 
can also result in the diminished prominence of spiral arms and the 
eventual transformation to an S0 galaxy (Larson, Tinsley, \& Caldwell 1980; 
Bekki, Couch, \& Shioya 2001, 2002; Conselice et al.\ 2001, 2003). 
Most spiral galaxies would use up their disk gas within a few Gyrs
(Gallagher, Bushouse, \& Hunter 1989). 

The paucity of spiral galaxies in the central regions of nearby
clusters, and the relative abundance of S0 galaxies relative to the
field population (e.g., Michard \& Marchal 1994), has been taken as
evidence of the transformation of spirals into lenticular galaxies
(e.g., Dressler et al.\ 1997).  Indeed,
observations at intermediate redshifts have revealed a higher
percentage of spirals in clusters than is observed locally.  
% (ref?) 
The presence of spiral galaxies in the periphery of massive 
clusters (Oemler 1974; Melnick \& Sargent 1977; Dressler
1980) has also been interpreted as testimony to their increased
survivability in this lower density environment.  But what about the
dwarf galaxies?

The lower gravitational potential of dwarf galaxies is suspected to 
make them particularly susceptible to ram-pressure stripping, although 
other factors, such as the velocity of the galaxy through the ICM and 
the density and temperature of the ICM, are also likely to be important.  
What ever the case, if the disruption process is not too severe, 
at least for the stars, it may leave behind the spiral structure. 
Once the gas is removed, spiral patterns are expected to disappear in 
less than 10 galactic rotations (Sellwood \& Carlberg 1984); which 
would suggest GMP~3629 was stripped within only the last couple of 
Gyrs.  The growing recognition of disks in intermediate-luminosity 
($-18\gtrsim M_B \gtrsim -20$) elliptical galaxies may be evidence 
of their increased ability to retain their disks, as compared with 
the fate of the lower-mass/lower-luminosity galaxies ($M_B\gtrsim -18$).  

% But is the absence of dwarf spiral galaxies and the large number of 
% dwarf elliptical galaxies in clusters evidence of dwarf galaxy 
% evolution.  For example, dSph and possibly dE galaxies may have 
% formed as they are; CDM models do predict the formation of small 
% primordial systems (e.g., Dekel \& Silk 1986; White \& Frenk 1991; 
% Bullock, Kravtsov, \& Weinberg 2000; Chiu, Gnedin, \& Ostriker 2001). 
% The smaller galaxies ($10^6$ -- $10^7$ M_{\sun}$) may 
% alternatively/additionally be the scattered fragments from major 
% mergers (e.g., Zwicky 1956; Gerola, Carnevali, \& Salpeter 1983; 
% Temporin et al.\ 2003) --- commonly referred to as tidal dwarf 
% galaxies.  Although Delgado-Donate et al.\ (2003) recently 
% concluded that, at  least in the field, galaxy-galaxy interactions 
% are not likely to produce many long-lived tidal dwarf galaxies. 

If, in addition to the dIrr galaxies observed in nearby clusters, 
dwarf spiral galaxies were to 
have entered and/or resided within the hazardous environment of a 
galaxy cluster, they too would be subject to the above mentioned 
processes and therefore likely be transformed into early-type 
galaxies.  But is there evidence that dwarf spirals were once a
predominant population in clusters? And if so, is there evidence
that they are the progenitors of today's cluster population of dwarf
ellipticals?

For over 25 years a substantial fraction of the galaxies in clusters 
at intermediate redshifts ($z>0.3$) were seen only as fuzzy blobs in 
ground-based images (Butcher \& Oemler 1978, 1984).  
HST images revealed that these objects are low luminosity 
spiral galaxies, often exhibiting disturbed morphologies (Couch et
al.\ 1994). Oemler et al.\ (1997) concluded that merging is an
implausible mechanism (see also Ostriker 1980) to explain the 
disturbed morphologies as the 
blue galaxy fraction is large and the merging probability is low. They
also noticed that disturbed spirals were observed throughout the
cluster. This would argue against ram pressure stripping 
being responsible for the disturbed spiral structure since 
this mechanism is expected to only operate efficiently near the cluster center. 
The ``galaxy harassment'' scenario (Moore et al.\ 1996),
however, does provide a plausible explanation to the disturbed morphology 
of the low luminosity spirals seen in intermediate cluster and their
transformation into dwarf early-type systems by the present
epoch.  Although direct mergers are extremely rare in the cluster
environment, every galaxy experiences a high speed close encounter
with a bright galaxy approximately once every Gyr. Moore et al.\ (1999)
show that these fly-by collisions have dramatic effects on the
morphologies of the dwarf galaxy population. The first encounter leads
to a pronounced bar instability. After several strong encounters, the
loss of angular momentum combined with impulsive heating leads to a
prolate shape supported equally by random motions and rotation. The
gas sinks to the very center of the galaxy and the stellar
distribution is heated to the extent that it closely resembles a dwarf
elliptical galaxy, although some may retain a very thick stellar disk
and would have the appearance of a dwarf lenticular.

The remarkably faint spiral/disk in GMP~3629 suggests that we may
indeed have an example of an incomplete morphological transformation
of a dwarf spiral into a dE or dS0 galaxy.  Although Jerjen et al.\
(2000b) stated that IC~3328 was not a dwarf spiral galaxy, the possible
detection of spiral arms in the dS0 galaxy FCC~288 (De Rijcke et al.\
2003), and the spiral patterns seen by Barazza et al.\ (2002) in three
Virgo dwarf galaxies is strong evidence for this scenario. Although 
our observations do not provide conclusive evidence in support
of the galaxy harassment model, there are additional hints in
our data that make this model particularly attractive. For instance,
there is the hint\footnote{The 
spiral arms in GMP~3629 do not appear 
to sprout from the galaxy center, but instead from the ends of 
a very faint bar --- although poor contrast makes this claim 
tentative.} 
of a bar-like structure in the residual images of GMP~3629. 
% as expected if this galaxy was at an early stage of morphological 
% transformation. 
We also remark that the large amount of
gas predicted to sink to the galaxy center may, in principle, provide
the fuel for forming the large stellar clusters seen in 
nucleated dwarf ellipticals.   

Due to the deficit of galaxies with spiral structure in dense
clusters, in such environments it is more common to think about 
how spiral density waves have been destroyed than to contemplate 
how spiral structure may form.  If created, 
such spiral patterns must be a transient phenomenon --- 
or at least they are destined to remain a faint feature --- otherwise 
they would have already been observed.  The bulge/disk decomposition in
Figure~\ref{fig4} suggests that most of GMP~3629's light is not from
the disk component, therefore ruling out the option that swing
amplification of noise, or clumps of material within the disk, 
may have caused the spiral-like pattern (Toomre 1981; Toomre \&
Kalnajs 1991).
Non-symmetric gravitational perturbations may invoke torques 
capable of generating spiral patterns in galaxy disks (e.g., 
Kormendy \& Norman 1979).  Rotating galactic bars 
% (e.g., Byrd et al.\ 1994), 
(e.g., Schwarz 1981), 
triaxial bulges (e.g. Trujillo et al.\ 2002), and/or 
triaxial dark halos (e.g., Bureau et al.\ 1999; 
Bekki \& Freeman 2002; Masset \& Bureau 2003) 
have been proposed as the culprit, at least when it comes 
to modifying the gas distribution. 
With regard to bars, while the gas can be forced about, 
Sellwood \& Sparke (1988) concluded that 
only the strongest of bars are likely to have a significant influence 
on the distribution of stars, a result confirmed by the structural 
analysis of bar and arm strength in Seigar \& James (1998; see 
also Seigar, Chorney, \& James 2003). 
Other mechanisms may be ``grooves'' in the distribution of the angular
momentum density (Sellwood \& Kahn 1991), or gravitational tides from
the passage of nearby objects (e.g., Toomre 1974 and references 
therein; Noguchi 1987; Sundelius et al.\ 1987).  At least in projection, 
all of the 18 galaxies reside close to 
luminous elliptical galaxies - indeed, all of these dwarf galaxies
reside in the same field as the HST pointings which were actually 
directed at luminous ellipticals
in the Coma cluster.  Simulations, however, have shown that such
``harassment'' from massive neighbors within a galaxy cluster
destroys, rather than creates, spiral patterns (Moore et al.\ 1996,
1998; Gnedin 2003).  

Lastly, we remark that 
intermediate mass black holes, having masses somewhere between those
of stellar-mass black holes and supermassive black holes, may have
formed in young compact star clusters via the runaway merging of
massive stars (Rees 1984; Kochanek, Shapiro \& Teukolsky 1987;
Matsushita et al.\ 2000; Ebisuzaki et al.\ 2001; Portegies Zwart \& 
McMillan 2002; Marconi et al.\ 2003). 
If so, one might
expect some of the dwarf ellipticals containing central star clusters
to harbor black holes.  On the other hand, perhaps the star clusters
are the result of the adiabatic growth of a pre-existing black hole
(e.g., van der Marel 1999).  If this is the case, then tests of the
adiabatic growth model that have excluded the nuclear component should
be reconsidered (e.g., Ravindranath, Ho, \& Filippenko 2002).

Clearly, more data are needed to provide 
conclusive evidence in support of or against the galaxy harassment
model. In particular, HST/STIS spatially-resolved spectroscopy of our
sample of Coma dEs may yield a critical test of this model by
measuring the amount of rotation and anisotropy in these galaxies, 
and also the ages of their nuclear clusters.

\acknowledgments 

We thank Bruno Binggeli and Agris Kalnajs for kindly reviewing this paper.
We also thank Chris Conselice and Jim Schombert whose comments helped
improve this work.  R.G.\ acknowledges funding from NASA grant
AR-08750.02-A and NRA-01-01-LTSA-080.\\ This research has made use of
the NASA/IPAC Extragalactic Database (NED) which is operated by the
Jet Propulsion Laboratory, California Institute of Technology, under
contract with the National Aeronautics and Space Administration.

\clearpage

\begin{figure}
\epsscale{0.75}
\plotone{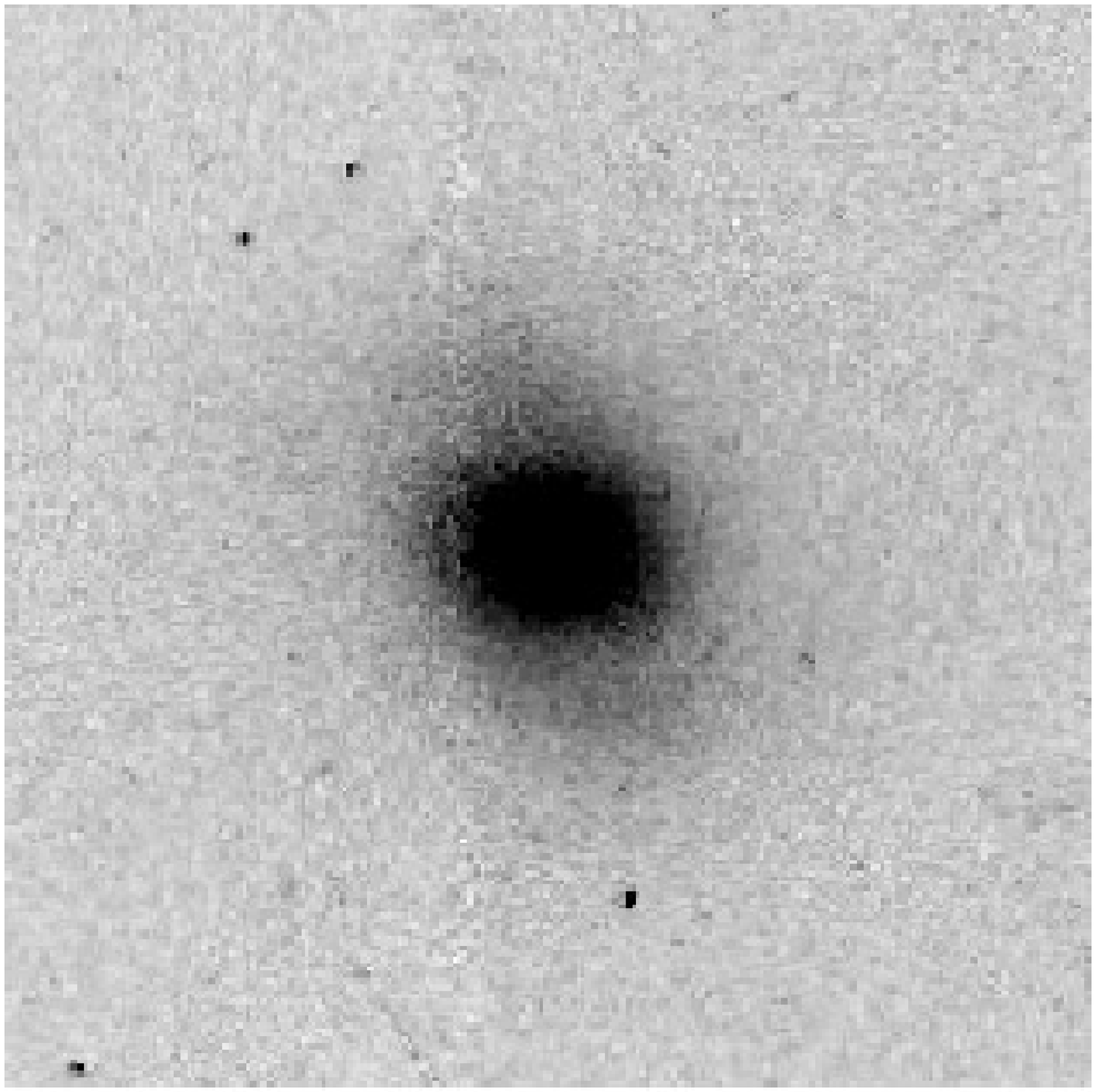}
\plotone{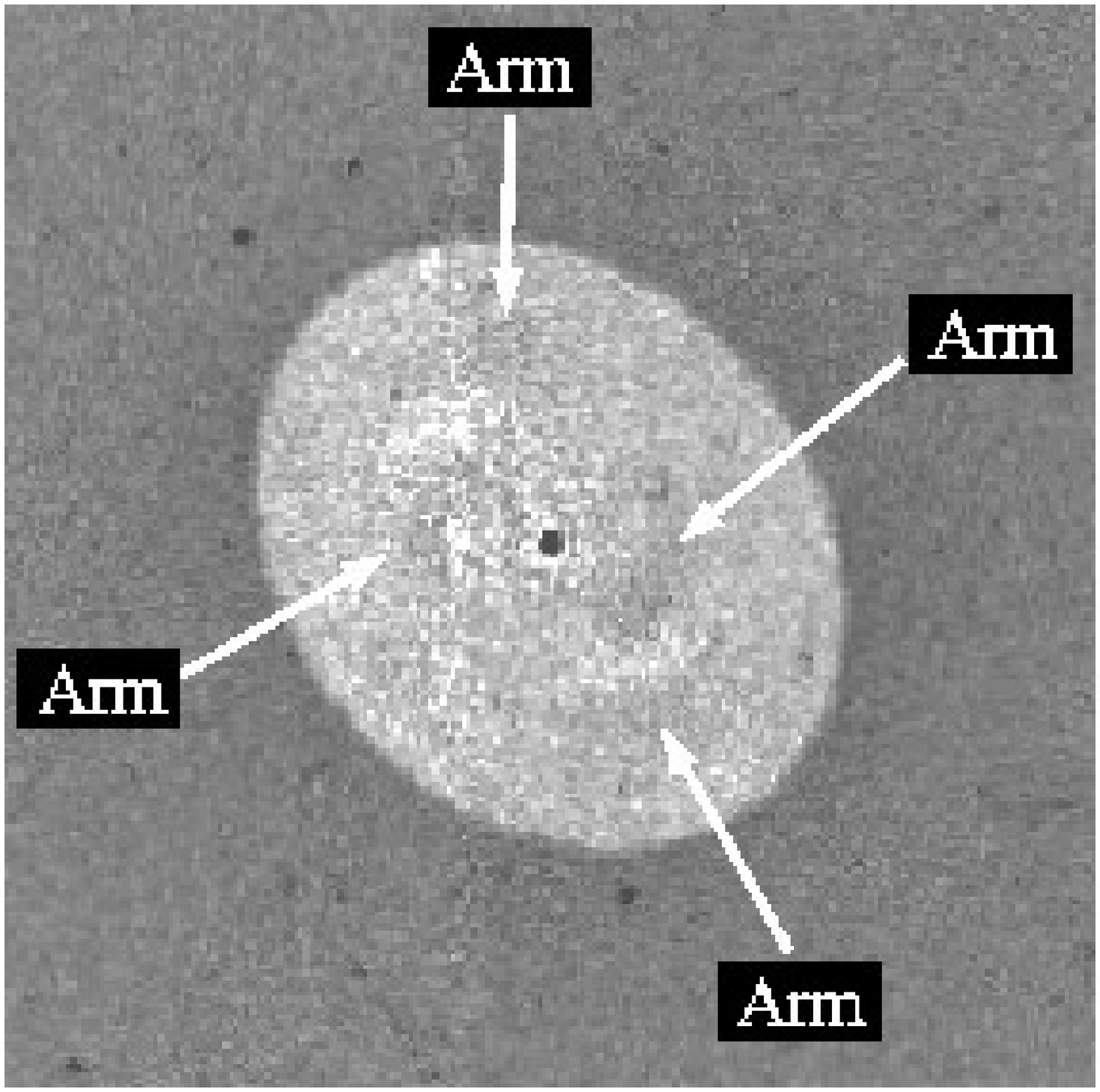}
\plotone{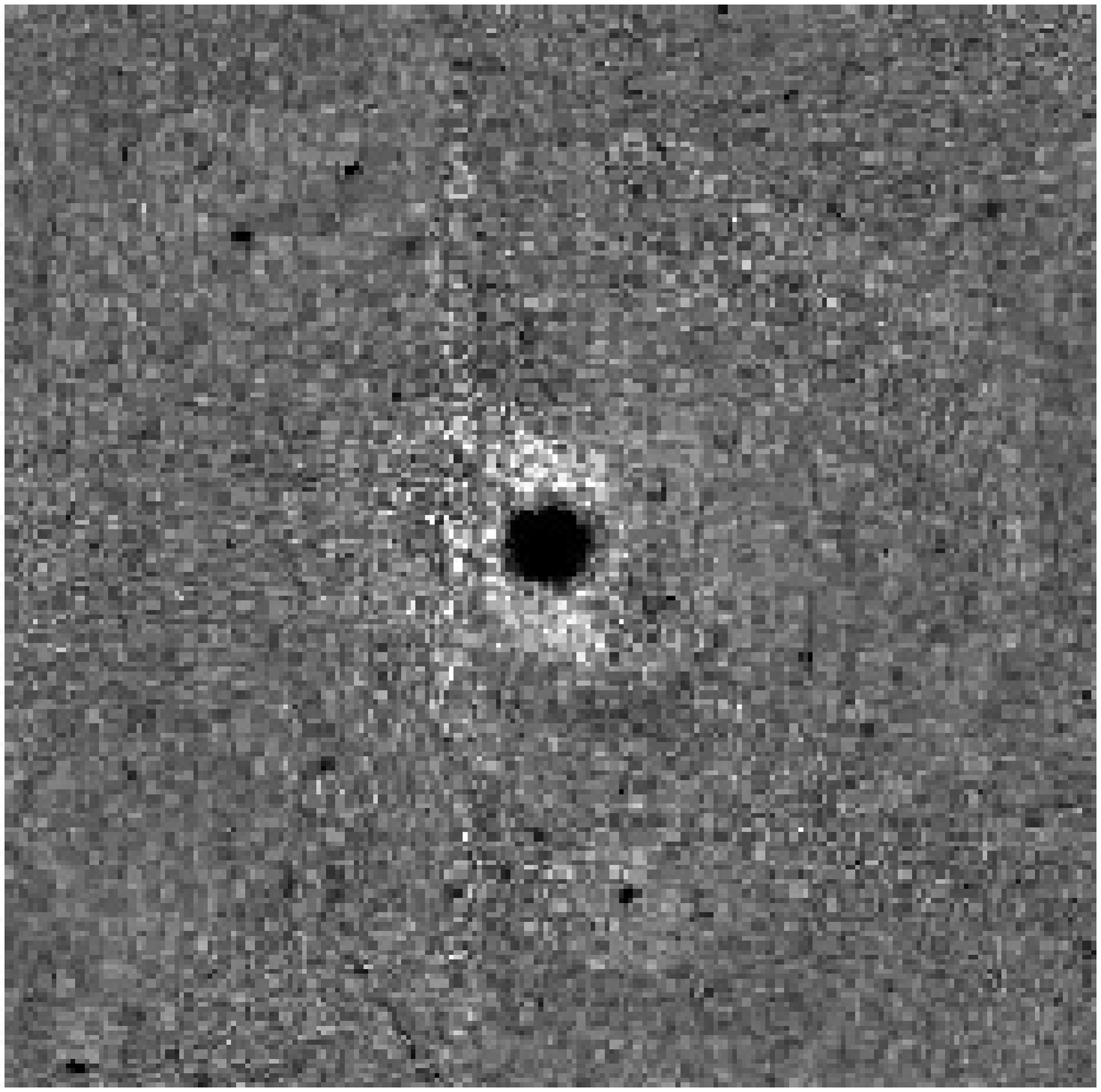}
\caption{
a) {\it HST} F606W image of GMP~3292.  b) Residual image of GMP~3292: 
following Jerjen et al.\ (2001) the axisymmetric component of the galaxy 
has been subtracted.  The labels ``arm'' are intended only to highlight
the visible portions of the arms, they do not necessarily imply that
there are 4 arms, there may well only be two (broken) arms.
c) Result of unsharp masking. 
% using a 7.5 pixel FWHM Gaussian smoothing filter. 
The image size is $20\arcsec\times20\arcsec$.
}
\label{fig1}
\end{figure}

\begin{figure}
\epsscale{0.75}
\plotone{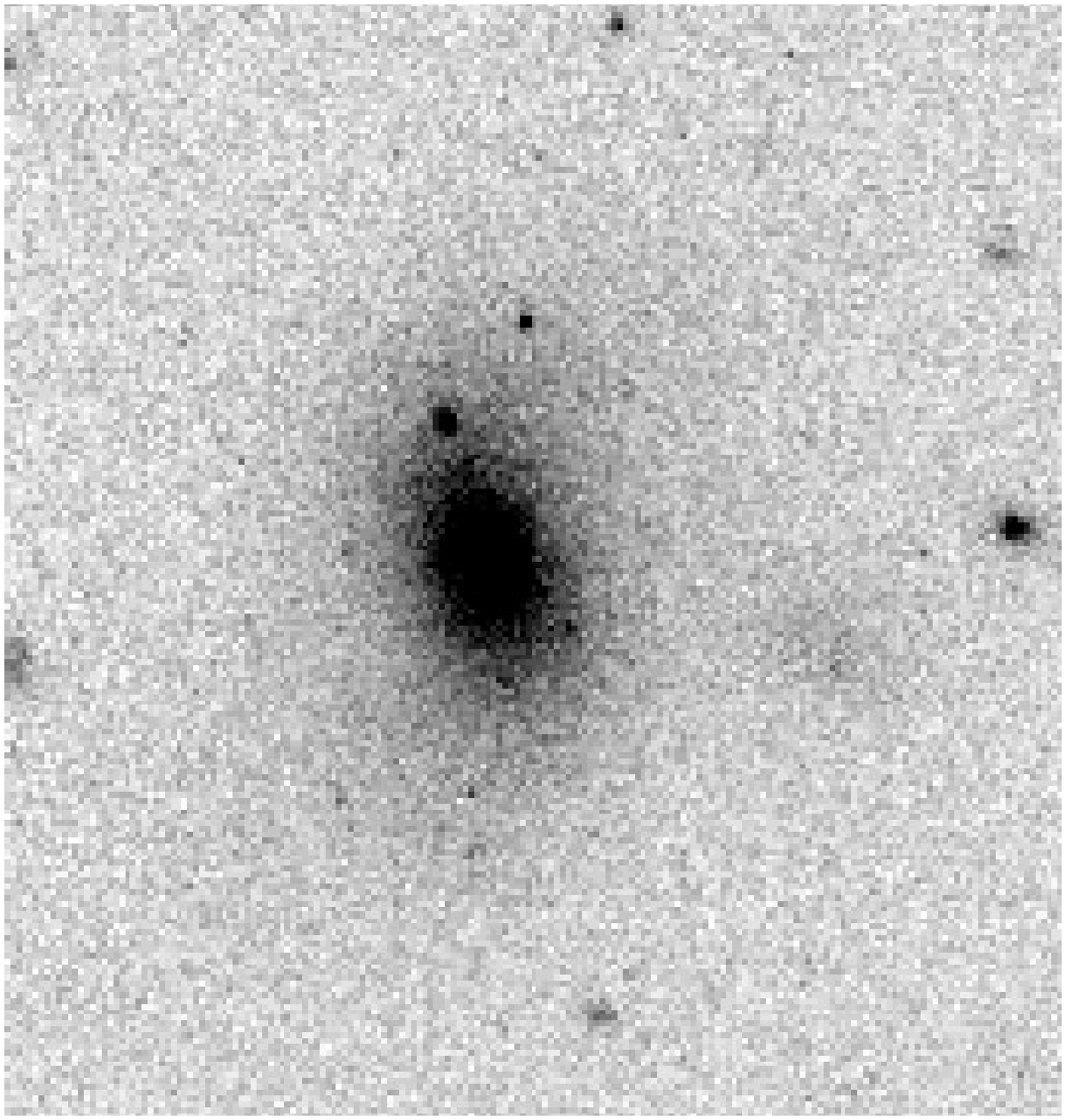}
\plotone{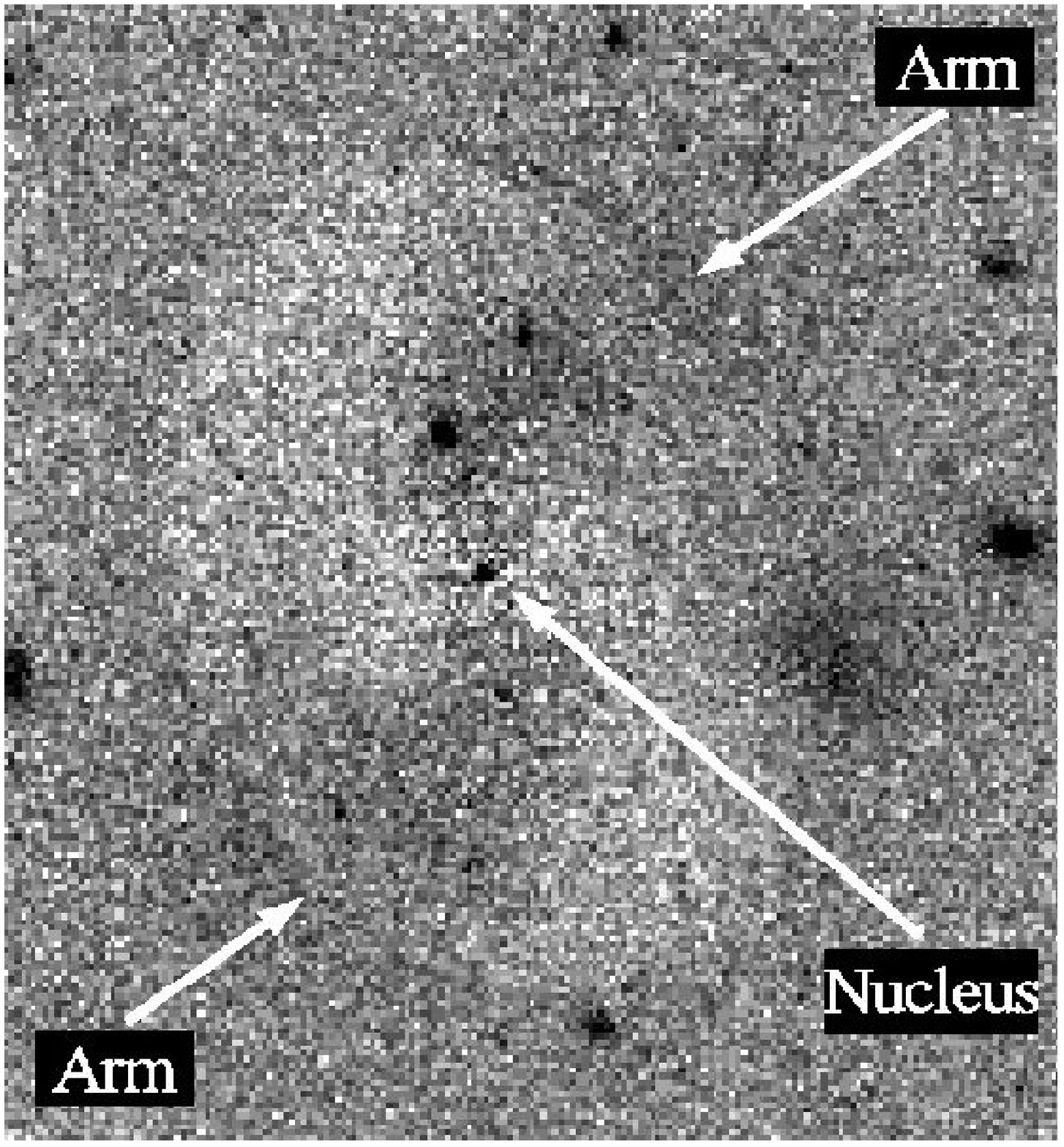}
\plotone{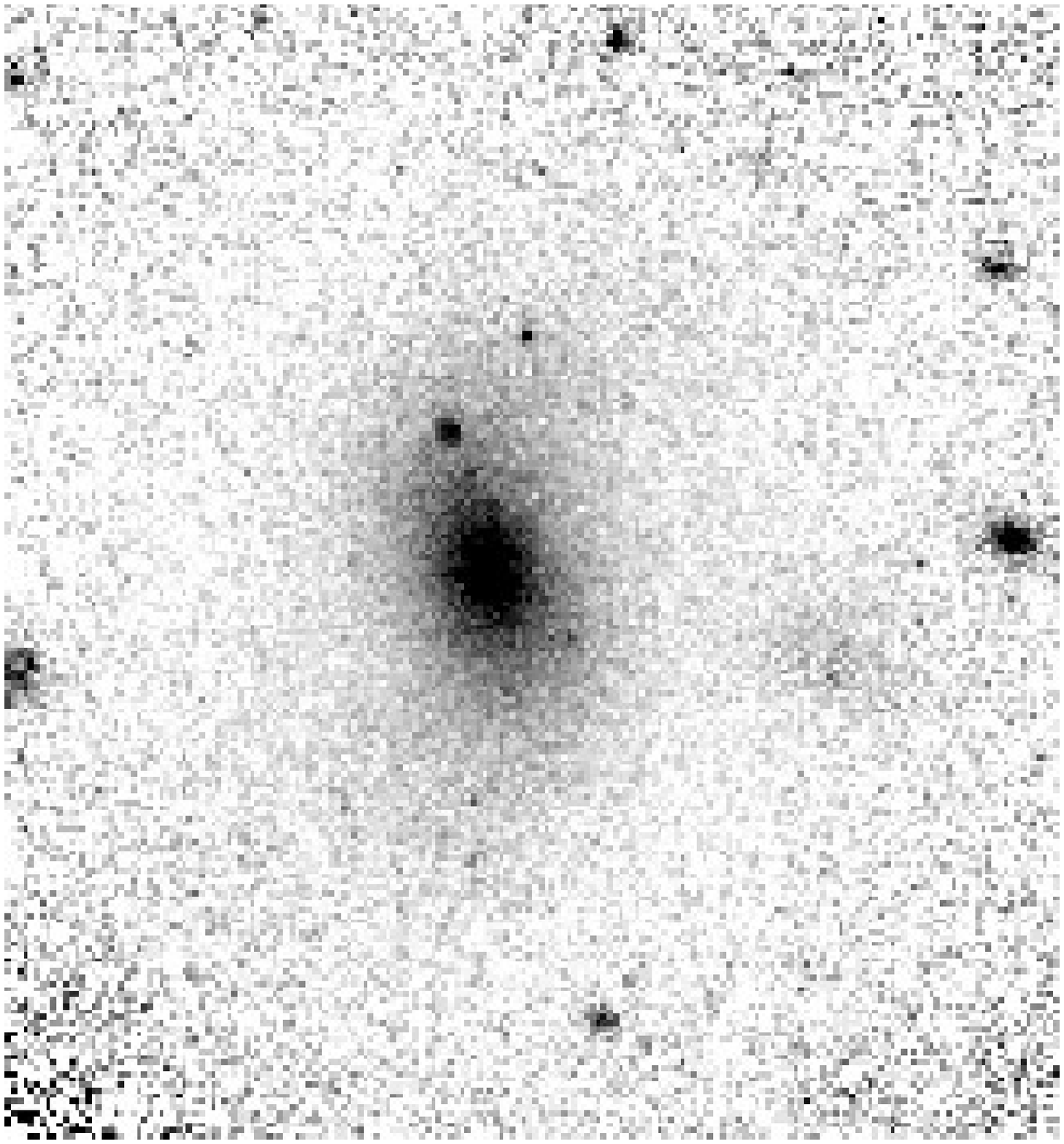}
\caption{Same as figure~\ref{fig1} but using GMP~3629.
The image size is $18.6\arcsec\times20\arcsec$.}
\label{fig2}
\end{figure}

\begin{figure}
\epsscale{0.75}
\plotone{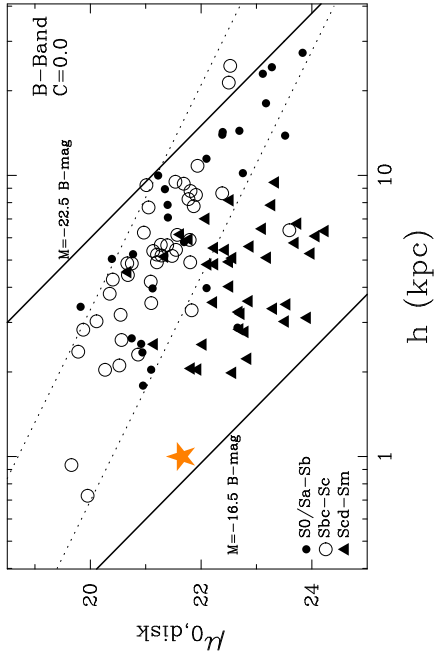}
\caption{Adaption of the $B$-band central disk surface brightnesses vs.\ 
disk scale-length diagram from Graham \& de Blok (2001). 
With the exception of GMP~3292 (star), all of the disks are in 
isolated, non-disturbed field galaxies. 
The solid lines are tracks of constant disk luminosity (slope = 5); for 
comparison, the dotted lines with a slope of 2.5 delineate the region 
where most early-type galaxies reside.  No inclination correction $C$ 
has been applied to the surface brightness terms.  GMP~3292's value of 
$\mu_0$ has been converted from $F606W$ to the $B$-band assuming a color 
of 1.08.  No change has been made to its scale-length. 
}
\label{fig3}
\end{figure}

\begin{figure}
\epsscale{0.75}
\plotone{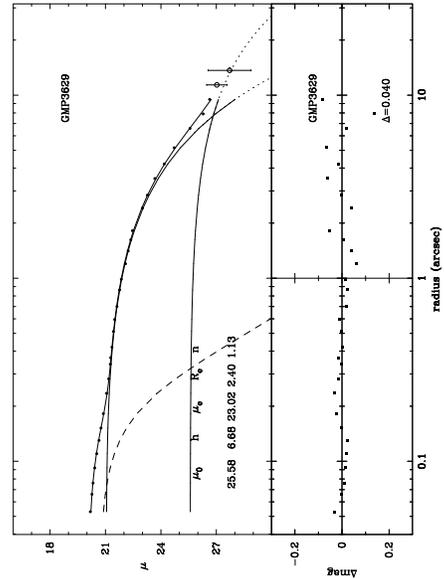}
\caption{Mean-axis ($r=\sqrt{ab}) F606W$ surface brightness profile of 
GMP~3629. 
The dashed line is a Moffat-convolved Gaussian fitted to the central
star cluster; the best-fitting, seeing-convolved bulge and disk models 
are shown with the solid curves.  All components were simultaneously fitted 
to the filled circles; the less reliable, low surface brightness data points
(denoted by the open circles) were not used in the fitting process.  
The extrapolation of the models are shown by the dotted lines. 
For comparison, a fit without a disk 
component is presented in Graham \& Guzm\'an (2003). 
}
\label{fig4}
\end{figure}

\clearpage
\begin{deluxetable}{lcccccc}
\tablecaption{Galaxy Data}
\tablewidth{0pt}
\tablehead{
\colhead{GMP}  & \colhead{$m_{F606W}$} & \colhead{$M_B$} & \colhead{B/D}  &
\colhead{$\mu_{0,F606W}$}  &  \colhead{$h_{F606W}$}  &  \colhead{$B-R$} \\
\colhead{No.}  & \colhead{mag} & \colhead{mag} & \colhead{ }  &
\colhead{mag arcsec$^{-2}$} & \colhead{kpc}  &  \colhead{mag}
}
\startdata
3292 & 16.74 & -17.28 & 0.3 & 20.7 & 1.0 & 1.5 \\ 
3629 & 18.02 & -16.00 & 2.6 & 25.6 & 3.1 & 1.4 \\
\enddata
\tablecomments{
Column 1: Godwin et al.\ 1983 (1983) catalog  number. 
Column 2: uncorrected apparent $F606W$-band magnitude, derived 
by extrapolating the fitted models to infinity. 
Column 3: Corresponding corrected (see text) absolute $B$-band magnitude. 
Column 4: Bulge-to-disk luminosity ratio.
Column 5 \& 6: uncorrected $F606W$-band central disk surface brightness 
and scale-length. 
Column 7: Global $B-R$ color term from Mobasher et al.\ (2001) and 
Guzm\'an (2003, priv.\ comm.). 
}
\end{deluxetable}

\end{document}